\documentclass[12pt]{article}

\usepackage{latexsym,amsmath,amssymb,amsthm,amsfonts,graphicx,natbib}
\usepackage{epsfig}
\usepackage{bm}

\include{def}
\def\dispmuskip{\thinmuskip= 3mu plus 0mu minus 2mu \medmuskip=  4mu plus 2mu minus 2mu \thickmuskip=5mu plus 5mu minus 2mu}
\def\textmuskip{\thinmuskip= 0mu                    \medmuskip=  1mu plus 1mu minus 1mu \thickmuskip=2mu plus 3mu minus 1mu}
\def\beq{\dispmuskip\begin{equation}}    \def\eeq{\end{equation}\textmuskip}
\def\beqn{\dispmuskip\begin{displaymath}}\def\eeqn{\end{displaymath}\textmuskip}
\def\bea{\dispmuskip\begin{eqnarray}}    \def\eea{\end{eqnarray}\textmuskip}
\def\bean{\dispmuskip\begin{eqnarray*}}  \def\eean{\end{eqnarray*}\textmuskip}

\def\paradot#1{\vspace{1.3ex plus 0.7ex minus 0.5ex}\noindent{\bf\boldmath{#1.}}}

\newcommand{\diag}{\text{diag}}

\newcommand{\wh}{\widehat}
\newcommand{\wt}{\widetilde}

\def\E{{\rm E}}                         

\def\I{1\!\!1} 				

\def\a{\alpha}

\def\s{\sigma}

\def\Sig{\Sigma}
\def\t{\theta}
\def\b{\beta}
\def\l{\lambda}

\def\N{{\cal N}}

\def\Kl{\text{\rm KL}}

\def\tr{\text{\rm tr}}

\def\cov{\text{\rm cov}}

\def\MSE{\text{\rm MSE}}

\def\diag{\text{\rm diag}}

\begin{document}

\title{Bayesian Adaptive Lasso with Variational Bayes for Variable Selection in High-dimensional Generalized Linear Mixed Models}
\author{
Dao Thanh Tung\thanks{Vietnam National University, Hanoi}
\and Minh-Ngoc Tran\thanks{The University of Sydney Business School. Correspondence to: minh-ngoc.tran@sydney.edu.au}
\and Tran Manh Cuong\footnotemark[1]
}
\maketitle

\begin{abstract}
This article describes a full Bayesian treatment for 
simultaneous fixed-effect selection and parameter estimation in high-dimensional generalized linear mixed models.
The approach consists of using a Bayesian adaptive Lasso penalty for signal-level adaptive shrinkage
and a fast Variational Bayes scheme for estimating the posterior mode of the coefficients.
The proposed approach offers several advantages over the existing methods,
for example, the adaptive shrinkage parameters are automatically incorporated, no Laplace approximation step is required to integrate out the random effects.
The performance of our approach is illustrated on several simulated and real data examples. 
The algorithm is implemented in the \texttt{R} package \texttt{glmmvb} and is made available online.

\bigskip
\noindent{\bf Keywords}: Posterior mode, Lasso, High dimensions, EM algorithm
\end{abstract}
\section{Introduction}\label{sec:Introduction}
Generalized linear mixed models (GLMMs) are widely used for modeling cluster-dependent data.
Variable selection in GLMMs is considered a difficult task, because of the present of integrals that are often analytically intractable.
Classical methods for variable selection, such as the ones based on hypothesis testing or subset selection,
are restricted to a few covariates.
Notable works are two recent papers by \cite{Groll:2012} and \cite{Schelldorfer:2013} which can do 
variable selection for GLMMs in high dimensions.
Their approach first estimates the likelihood by approximating the integrals over the random effects using the Laplace method,
then minimizes the sum of this estimated likelihood and a Lasso-type penalty which is the $l_1$-norm of the fixed effect coefficients.
Using a Lasso-type penalty will shrink the coefficients towards zero, thus leading to variable selection. 
This variable selection approach is attractive compared to the classical approaches as it can handle problems with a large number of potential covariates.

However, there is still room for improvement within the approach of \cite{Groll:2012} and \cite{Schelldorfer:2013}.
First, the Laplace approximation of the likelihood might be in some cases not very accurate \citep[see, e.g.][]{Joe:2008}.
Second, the performance depends on the shrinkage parameter that needs to be selected appropriately.
So that the user has to run the procedure over and over again for different values of the shrinkage parameter   
within a pre-specified range, then selects the best value of the shrinkage parameter based on some criterion such as AIC or BIC.
As the result, the entire procedure for selecting the final model may be time consuming.
Furthermore, specifying an appropriate range for the shrinkage parameter is not straightforward.
Third, this approach uses the same shrinkage parameter for every coefficients,
which can lead to biased estimates of the coefficients.

This article proposes using the Bayesian adaptive Lasso for variable selection in high-dimensional GLMMs.
We use double exponential priors for the coefficients with different shrinkage parameters for different coefficients,
which is equivalent to the approach in \cite{Groll:2012} and \cite{Schelldorfer:2013} when all the shrinkage parameters are equal.
It is desirable to apply different shrinkage on different coefficients
to achieve adaptivity,
i.e. larger shrinkage should be put on coefficients corresponding to unimportant covariates and 
smaller shrinkage should be used for important covariates \citep{Zou:2006}.
We consider a full Bayesian treatment, i.e. we put appropriate priors on all the model parameters,
including the shrinkage parameters. 
As the result, we overcome the challenging task of selecting a high-dimensional vector of the shrinkage parameters.

We then develop a variational Bayes (VB) algorithm for estimating the posterior mode of the coefficient vector
and the posterior distribution of the covariance matrix of the random effects.
This leads to a totally automatic procedure for simultaneous variable selection and parameter estimation in GLMMs,
and the adaptive shrinkage parameters are automatically incorporated. 
Finally, unlike the approach in \cite{Groll:2012} and \cite{Schelldorfer:2013}, our approach does not rely on the Laplace approximation for integrating out the random effects,
because the updating procedure in the variational Bayes algorithm leads to an integral that either can be computed analytically
or approximated in close form with an arbitrary accuracy.
The examples in Section \ref{sec:examples} show that our approach outperforms the existing methods
in terms of the rate of correctly-fitted models, the mean squared error of the estimates, and the CPU running time.

The paper is organized as follows.
Section \ref{sec:VB theory} provides some background on the variational Bayes method,
and presents the VB method for estimating the posterior mode.
Section \ref{sec:vbglmm} describes our algorithm for variable selection in GLMMs.
Section \ref{sec:examples} presents a systematic simulation example and real data applications.
Section \ref{sec:conclusion} concludes and discusses some possible extensions.
The algorithm is implemented in the \texttt{R} package \texttt{glmmvb} and is available at \texttt{https://sites.google.com/site/mntran26/research}.

\section{Variational Bayes method}\label{sec:VB theory}
Suppose we have data $y$, a likelihood $p(y|\theta)$ where $\theta\in\mathbb{R}^d$ is an unknown parameter, and a prior distribution $p(\theta)$ for $\theta$. 
Variational Bayes (VB) approximates the posterior $p(\t|y)\propto p(\t)p(y|\t)$ by a distribution $q(\t)$
within some more tractable class, chosen to minimize the Kullback-Leibler divergence
\beq\label{eq:KL_distance}
\Kl(q\|p) = \int q(\t)\log\frac{q(\t)}{p(\t|y)}d\t.
\eeq
We have 
\beqn
\log p(y)=\int q(\t)\log\frac{p(y,\t)}{q(\t)}d\t+\int q(\t)\log\frac{q(\t)}{p(\t|y)}d\t=L(q)+\Kl(q\|p),
\eeqn
where
\beq\label{eq:original VB}
L(q) = \int q(\t)\log\frac{p(y,\t)}{q(\t)}d\t.
\eeq
As $\Kl(q\|p)\geq0$, $\log\;p(y)\geq L(q)$ for every $q(\t)$. $L(q)$ is therefore often called the lower bound,
and minimizing $\Kl(q\|p)$ is equivalent to maximizing $L(q)$.

Often factorized approximations to the posterior are considered in variational Bayes.  We explain the idea for a factorization with 2 blocks.
Assume that $\t=(\t_1,\t_2)$ and that $q(\t)$ is factorized as 
\beq\label{eq:VBfull}
q(\t)=q_1(\t_1)q_2(\t_2).
\eeq
We further assume that $q_1(\t_1)=q_{\tau_1}(\t_1)$ and $q_2(\t_2)=q_{\tau_2}(\t_2)$ where $\tau_1$ and $\tau_2$ are variational parameters that need to be estimated.
Then
\bean
L(\tau_1,\tau_2)=L(q)&=&\int q_{\tau_1}(\t_1)q_{\tau_2}(\t_2)\log p(y,\t)d\t_1d\t_2-\int q_{\tau_1}(\t_1)\log q_{\tau_1}(\t_1)d\t_1+C(\tau_2)\\
&=&\int q_{\tau_1}(\t_1)\left(\int q_{\tau_2}(\t_2)\log p(y,\t)d\t_2\right)d\t_1-\int q_{\tau_1}(\t_1)\log q_{\tau_1}(\t_1)d\t_1+C(\tau_2)\\
&=&\int q_{\tau_1}(\t_1)\log\wt p(y,\t_1)d\t_1-\int q_{\tau_1}(\t_1)\log q_{\tau_1}(\t_1)d\t_1+C(\tau_2)\\
&=&\int q_{\tau_1}(\t_1)\log\frac{\wt p_1(y,\t_1)}{q_{\tau_1}(\t_1)}d\t_1+C(\tau_2),
\eean
where $C(\tau_2)$ is a constant depending only on $\tau_2$ and 
\beqn
\wt p_1(y,\t_1)=\exp\left(\int q_{\tau_2}(\t_2)\log p(y,\t)d\t_2\right)=\exp\big(\E_{-\t_1}(\log p(y,\t))\big).
\eeqn
Given that $\tau_2$ is fixed. Let
\beq\label{eq:lam1}
\tau_1^* = \tau_1^*(\tau_2) = \arg\max_{\tau_1}\left\{\int q_{\tau_1}(\t_1)\log\frac{\wt p_1(y,\t_1)}{q_{\tau_1}(\t_1)}d\t_1\right\},
\eeq
then 
\beq\label{eq:L_increase1} 
L(\tau_1^*,\tau_2)\geq L(\tau_1,\tau_2)\;\;\text{for all}\; \tau_1.
\eeq
Similarly, given a fixed $\tau_1$, let
\beq\label{eq:lam2}
\tau_2^* = \tau_2^*(\tau_1) = \arg\max_{\tau_2}\left\{\int q_{\tau_2}(\t_2)\log\frac{\wt p_2(y,\t_2)}{q_{\tau_2}(\t_2)}d\t_2\right\},
\eeq
with
\beqn
\wt p_2(y,\t_2)=\exp\left(\int q_{\tau_1}(\t_1)\log p(y,\t)d\t_1\right)=\exp\big(\E_{-\t_2}(\log p(y,\t))\big).
\eeqn
Then,
\beq\label{eq:L_increase2} 
L(\tau_1,\tau_2^*)\geq L(\tau_1,\tau_2)\;\;\text{for all}\; \tau_2.
\eeq
Let $\tau^\text{old}=(\tau^\text{old}_1,\tau^\text{old}_2)$ be the current value of $\tau_1$ and $\tau_2$,
update $\tau^\text{new}_1=\tau_1^*(\tau^\text{old}_2)$ as in \eqref{eq:lam1}
and $\tau^\text{new}_2=\tau_2^*(\tau^\text{new}_1)$ as in \eqref{eq:lam2}. 
Then, because of \eqref{eq:L_increase1} and \eqref{eq:L_increase2},  
\beq\label{eq:lb_increase}
L(\tau^\text{new})\geq L(\tau^\text{old}).
\eeq
This leads to an iterative scheme for updating $\tau$
and \eqref{eq:lb_increase} ensures the improvement of the lower bound over the iterations.
Because the lower bound $L(\tau)$ is bounded from above by $\log p(y)$, the convergence of the iterative scheme is guaranteed.
The above argument can be easily extended to the general case in which $q(\theta)$ is factorized into $K$ blocks $q(\t)=q_1(\t_1)\times...\times q_K(\t_K)$.

The variational Bayes approximation is now reduced to solving an optimization problem in the form of \eqref{eq:lam1}.
Let $\wt p_1(\t_1|y)$ be the density of $\t_1$ determined by the unnormalized function $\wt p_1(y,\t_1)$, i.e.
\beq\label{eq:optimal_VB}
\wt p_1(\t_1|y)=\frac{\wt p_1(y,\t_1)}{\int \wt p_1(y,\t_1)d\t_1}\propto   \exp\big(\E_{-\t_1}(\log p(y,\t))\big).
\eeq
In many cases, a conjugate prior $p(\t_1)$ can be selected such that $\wt p_1(\t_1|y)$ belongs to a family of recognizable parametric densities. 
Then the optimal VB posterior $q_{\tau_1^*}(\t_1)$ that maximizes the integral on the right hand side of \eqref{eq:lam1} is 
$\wt p_1(\t_1|y)$, with $\tau_1^*$ the corresponding parameter of this density.

If $\wt p_1(\t_1|y)$ does not belong to a recognizable density family,
some optimization technique is needed to solve \eqref{eq:lam1}.
Note that \eqref{eq:lam1} has exactly the same form as the original VB problem that attempts to maximize $L(q)$ in \eqref{eq:original VB}.
We can first select a functional form for the variational distribution $q$ and then estimate the 
unknown parameters accordingly. 
If the variational distribution is assumed to belong to the exponential family with unknown parameters $\tau$,
\cite{Salimans:2013} propose a stochastic approximation method for solving for $\tau$.
The reader is referred to their paper for the details.

\subsection{Variational Bayes method for estimating the posterior mode}\label{subsec:VB theory}
As pointed out in \cite{Tibshirani:96}, the Lasso estimator 
is equivalent to the posterior mode when a double-exponential prior (also called Laplace prior) is used for the vector of coefficients $\beta$.
In general, for the variable selection purposes in Bayesian settings,
one is interested in the posterior mode rather than the entire posterior distribution.
As will be seen in the next section, variable selection in GLMMs
is carried out through computing the posterior mode of the fixed-effect coefficient vector $\beta$.
We will present in this section a Variational Bayes method for estimating a posterior mode.

Write $\t=(\t_1,\t_2)$,
where $\t_1$ is the vector of parameters whose posterior mode is of our
interest, and $\t_2$ is a vector of other parameters, random effects or missing data.
Then, we can use a VB posterior of the form 
\beq\label{eq:VBmode}
q(\t)=\delta_{\tau_1}(\t_1)q_{\tau_2}(\t_2),
\eeq
with $\delta_{\tau_1}(\t_1)$ a point mass density concentrated at $\tau_1$.
For our purposes, $\tau_1$ will be the estimate
of the posterior mode of $\t_1$.

Equations \eqref{eq:lam1} and \eqref{eq:lam2} become
\beq\label{eq:lam1'}
\tau_1^*(\tau_2) = \arg\max_{\tau_1}\int q_{\tau_2}(\t_2)\log p(y,\tau_1,\t_2)d\t_2\tag{\ref{eq:lam1}'},
\eeq
and 
\beq\label{eq:lam2'}
\tau_2^*(\tau_1) = \arg\max_{\tau_2}\left\{\int q_{\tau_2}(\t_2)\log\frac{p(y,\tau_1,\t_2)}{q_{\tau_2}(\t_2)}d\t_2\right\}\tag{\ref{eq:lam2}'}.
\eeq
The optimal VB posterior of $\t_2$ from \eqref{eq:lam2'} is $q_{\tau_2^*}(\t_2)=p(\t_2|y,\tau_1)\propto p(y,\tau_1,\t_2)$.
Then \eqref{eq:lam1'} and \eqref{eq:lam2'} can be written in terms of the EM algorithm \citep{Dempster:1977}, where  
\begin{itemize}
\item E-step: compute $Q(\tau_1|\tau_1^\text{old})=\int p(\t_2|y,\tau_1^\text{old})\log p(y,\tau_1,\t_2)d\t_2$.
\item M-step: maximize $Q(\tau_1|\tau_1^\text{old})$ over $\tau_1$.
\end{itemize}
The EM algorithm therefore can be considered as a special case of this VB algorithm where
$q_{\tau_2}(\t_2)$ in \eqref{eq:VBmode} is $q_{\tau_2}(\t_2)=p(\t_2|y,\tau_1)$.
Note that the VB mode method in \eqref{eq:lam1'} and \eqref{eq:lam2'}
is somewhat more flexible than the EM algorithm because 
we have more freedom to find a solution to \eqref{eq:lam2'} provided that $q_{\tau_2}(\t_2)$ is restricted to some density family. This is important because the optimal density $q_{\tau_2^*}(\t_2)=p(\t_2|y,\tau_1)$ in some cases does not belong to a family of recognizable densities,
and it is difficult to compute the integral in the E-step. 
For example, in generalized linear mixed models considered in this paper,
the distribution of the random effects conditional on the data and the other parameters
does not belong to a family of recognizable densities,
making it difficult to estimate the coefficient vector using the EM algorithm.

\section{Variable selection and estimation for GLMMs}\label{sec:vbglmm}
Consider a generalized linear mixed model 
where $y_i=(y_{i1},...,y_{in_i})'$ is the vector of responses for the $i$th subject, $i=1,...,m$.
Given random effects $b_i$, the $y_{ij}$ are
conditionally independently distributed with the density or probability function
\beqn
f(y_{ij}|\beta,b_i)=\exp\left(\frac{y_{ij}\eta_{ij}-\zeta(\eta_{ij})}{\phi}+c(y_{ij},\phi)\right),
\eeqn
where $\eta_{ij}$ is a canonical parameter which is monotonically related to the conditional
mean $\mu_{ij}=E(y_{ij}|\beta,b_i)$ through a link function $g(\cdot)$, $g(\mu_{ij})=\eta_{ij}$.
The fixed effect coefficient vector is $\beta=(\b_0,\b_{1:p}')'$ with 
$\b_0$ the slope and $\b_{1:p}=(\b_1,...,\b_p)'$.
The scale parameter $\phi$ can be unknown and $\zeta(\cdot)$ and $c(\cdot)$ are known functions.  
Here, for simplicity, we are considering the case of a canonical link function,
i.e. $g(\mu_{ij})=\eta_{ij}$.  
The vector $\eta_i=(\eta_{i1},...,\eta_{in_i})'$ is modeled as
$\eta_i=\b_01_{n_i}+X_i\beta_{1:p}+Z_i b_i$,
where $1_{n_i}$ is the vector of ones,
$X_i$ is an $n_i\times p$ design matrix for the
fixed effects and $Z_i$ is an $n_i\times u$ design matrix for the random effects (where $u$ is the dimension
of $b_i$).  Let $n=\sum_{i=1}^mn_i$, $b=(b_1',...,b_m')'$ and
\beqn
y = \begin{pmatrix}
y_1\\
y_2\\
\vdots\\
y_m
\end{pmatrix},\;\;
X = \begin{pmatrix}
1& X_1\\
1& X_2\\
\vdots&\vdots\\
1& X_m
\end{pmatrix},\;\;Z=\begin{pmatrix}
Z_1&0&\cdots&0\\
0&Z_2&\cdots&0\\
\vdots&\vdots&\cdots&\vdots\\
0&0&\cdots&Z_m
\end{pmatrix},\;\;\eta = \begin{pmatrix}
\eta_1\\
\eta_2\\
\vdots\\
\eta_m
\end{pmatrix}=X\beta+Zb.
\eeqn
The likelihood conditional on the random effects $b$ is
\beqn
p(y|\beta,b,\phi)=\prod_{i=1}^m\prod_{j=1}^{n_i}f(y_{ij}|\beta,b_i)=\exp\left(\frac{1}{\phi}(y'\eta-1'\zeta(\eta))+c(y,\phi)\right),
\eeqn
where $\zeta(\eta)$ is understood component-wise and $c(y,\phi)=\sum_{i,j}c(y_{ij},\phi)$.

The random effects $b_i$ are often assumed independently distributed as $\N(0,Q^{-1})$,
where $\N(\mu,\Sigma)$ denotes the multivariate normal distribution with mean $\mu$ and covariance matrix $\Sigma$.
The distribution of $b$ is $\N(0,Q^{-1}_b)$ with $Q_b$ a block diagonal matrix $\text{diag}(Q,...,Q)$.
We consider Bayesian inference with the following hierarchy
\bea\label{eq:BaLasso_GLMM}
y|\beta,b,\phi&\sim&p(y|\beta,b,\phi)\\
b|Q&\sim& \N(0,Q^{-1}_b)\notag\\
Q&\sim&\text{Wishart}(S_0,\nu_0)\notag\\
p(\b_0)&\sim&1\notag\\ 
\beta_j|\l_j&\sim&\text{DE}(\l_j)=\frac{\l_j}{2}\exp(-\l_j|\b_j|),\ j=1,...,p\notag\\
\lambda_j&\sim&\text{Gamma}(r,s) =\frac{s^r}{\Gamma(r)}(\l_j)^{r-1}\exp(-s\l_j)\notag,
\eea
where $\text{DE}(\l_j)$ denotes the double-exponential density.
If $\phi$ is unknown we also put a prior $p(\phi)$ on $\phi$.
We refer to the suggested model \eqref{eq:BaLasso_GLMM} as the Bayesian adaptive Lasso model (BaLasso) for GLMM.
The set of model parameters is $\theta=(\beta,Q,\phi,b,\l_1,...,\l_p)$ and
$S_0,\nu_o,r,s$ are hyperparameters whose selection is discussed later.

When $\l_j=\l$ and considered fixed, 
the joint posterior distribution of $\b,Q,\phi$ is
\beqn
p(\b,Q,\phi)\propto p(\phi)p(Q)\exp\left(\log\int p(y|\b,b,\phi)p(b|Q)db-\l\sum_{j=1}^p|\b_j|\right).
\eeqn
In this case, the posterior marginal mode of $\b$ from model \eqref{eq:BaLasso_GLMM} is exactly
the penalized maximum likelihood estimate in \cite{Groll:2012} and \cite{Schelldorfer:2013},
who estimate the parameters by maximizing 
\beq\label{eq:Groll&Schelldorfer}
\log\int p(y|\b,b,\phi)p(b|Q)db-\l\sum_{j=1}^p|\b_j|
\eeq
over $\b$.
Note that we use different $\l_j$ for different coefficient $\b_j$ to achieve signal-level adaptivity \citep{Zou:2006}. 

The Bayesian Lasso was first proposed in \cite{Park:2008} who considered a single shrinkage $\l$ for all coefficients,
in the context of ordinary linear regression only.
The Bayesian adaptive Lasso for GLMs was proposed in \cite{Griffin:2011} and \cite{Leng:2013}.
\cite{Griffin:2011} employed the EM algorithm to estimate the posterior mode of $\beta$
and were therefore able to carry out variable selection.
\cite{Leng:2013} first used Gibbs sampling to sample from the posterior of $\l$ and then proposed a Bayesian-frequentist hybrid method for doing variable selection where $\l$ is fixed to its posterior mode.
To the best of our knowledge, this paper is the first to consider the Bayesian adaptive Lasso model \eqref{eq:BaLasso_GLMM}
for inference in GLMMs, and also use VB for estimating a posterior mode.

We use Variational Bayes to approximate the posterior $p(\t|y)$ with the variational posterior factorized as 
\beq\label{VB factorization GLMM}
q(\t)=q(\beta)q(Q)q(\phi)q(b)\prod_{j=1}^p q(\l_j)
\eeq
where $q(\b)=\delta_{\beta^q}(\b)$ and 
$q(b)$ is normal with mean $\mu_b^q$ and covariance matrix $\Sigma_b^q$. 
From \eqref{eq:lam1'}, the mode estimate $\b^q$ is updated by
\beq\label{eq:beta update glmm}
\b^q=\arg\max_\b\left\{\exp\left(\E_{-\b}(\log p(y,\theta))\right)\right\}=\arg\max_\b\left\{[\frac{1}{\phi}]\int \big(y'\eta-1'\zeta(\eta))\big)q(b)db-\sum_{j=1}^p[\l_j]|\b_j|\right\}.
\eeq
Hereafter, $[\cdot]$ denotes the expectation with respect to the VB posterior.
Solving this optimization problem is discussed in detail later on.

For the normal linear mixed regression model,
the optimal VB posterior $q(b)$ is a normal distribution
and therefore the parameters $\mu_{b}^q,\Sig_{b}^q$ are updated in closed form.
In the other cases, from \eqref{eq:optimal_VB}, the optimal VB approximation $q(b)$ is
\beq
q(b)\propto\exp\left(-\frac12b'[Q_b]b+[\frac{1}{\phi}](y'\eta-1'\zeta(\eta))\right)
\eeq
with $\eta=X\beta^q+Zb$. This distribution does not have the form of a standard distribution.
We suggest using the Gaussian approximation to approximate this optimal distribution by a normal distribution with mean 
$\mu_{b}^q$ and covariance matrix $\Sig_{b}^q$.
Let $b^*$ be the maximizer of the function
\[h(b)=-\frac12b'[Q_b]b+[\frac{1}{\phi}](y'\eta-1'\zeta(\eta)),\]
which can be easily found by the Newton-Raphson method (see Appendix C). 
Then, $\mu_{b}^q$ and $\Sig_{b}^q$ are updated as follows
\bea\label{eq:update mu_b}
\mu_{b}^q &=& b^*\notag\\
\Sig_{b}^q &=&\left( [\frac{1}{\phi}]Z' \mbox{diag}\left(\ddot{\zeta}(\eta^*)\right)Z + [Q_b]\right)^{-1}.
\eea
with $\eta^*=X\beta^q+Zb^*$.

The optimal VB posterior $q(Q)$ is a Wishart with degrees of freedom and scale matrix
\beq\label{eq:update Q}
\nu^q=\nu_0+m,\;\;S^q=\left(S_0^{-1}+\sum_{i=1}^m(\mu_{b_i}^q{\mu_{b_i}^q}'+\Sig_{b_i}^q)\right)^{-1},
\eeq
where $\mu_{b_i}^q$ and $\Sig_{b_i}^q$ are extracted from $\mu_{b}^q$ and $\Sig_{b}^q$ accordingly. Then, $[Q_b]=\text{diag}([Q],...,[Q])$ with $[Q]=\nu^qS^q$.

The optimal VB posterior of $\l_j$ is Gamma with shape and rate
\beq\label{eq:update lambda}
\a^q_{\l_j}=r+1,\;\;\;\b^q_{\l_j}=|\b_j^q|+s,
\eeq
and therefore $[\l_j]={\a_{\l_j}^q}/{\b_{\l_j}^q}$.
In many cases such as Poisson and logistic regression, $\phi$ is a known constant,
otherwise we can put a suitable prior on $\phi$ such that the optimal
\beq\label{eq:update_phi}
q(\phi)\propto \exp\big(\E_{-\phi}(\log p(y,\theta))\big)
\eeq
belongs to a recognizable family.
In the case of normal linear mixed regression, for example, if using an inverse Gamma prior 
with shape $\a_{\s^2}^0$ and scale $\b_{\s^2}^0$ for the dispersion parameter $\phi=\s^2$,
the optimal VB posterior $q(\s^2)$ is an inverse Gamma with shape and scale
\beqn\label{eq:update_phi normal}
\a_{\s^2}^q=n/2+\a_{\s^2}^0,\;\;\; \b_{\s^2}^q=\frac12\|y-X\b^q-Z\mu_b^q\|^2+\frac12\tr(Z\Sig_b^qZ')+\b_{\s^2}^0.
\eeqn
In this case, $[1/\s^2] = {\a_{\s^2}^q}/{\b_{\s^2}^q}$.

We summarize below the VB algorithm for doing variable selection in GLMMs.

\paradot{VBGLMM algorithm}
\begin{enumerate}
\item Initialize $\beta^q$ and $S^q$ (and $q(\phi)$ if applicable).
\item Update $\a_{\l_j}^q$ and $\b_{\l_j}^q$ as in \eqref{eq:update lambda}.
\item Update $\mu_{b}^q$ and $\Sig_{b}^q$ as in \eqref{eq:update mu_b}
\item Update $S^q$ as in \eqref{eq:update Q}.
\item Update $\beta^q$ as in \eqref{eq:beta update glmm}.
\item Update $q(\phi)$ (if applicable).
\item Repeat Steps 2-6 until convergence.
\end{enumerate}

We may initialize $\b^q$ to some initial estimate such as the MLE if available.
We suggest to stop the iteration when the difference between two successive updates of the main parameters $\b^q$ is smaller than some prespecified value. 

\paradot{Selection of the hyperparameters}
For the prior on the $\l_j$, one can use the improper scale-invariant prior $p(\l_j)\propto 1/\l_j$, i.e. $r=s=0$.
In this paper, we use the empirical Bayes method
as in \cite{Park:2008} and \cite{Leng:2013} for selecting $r$.
We use a Gamma prior, Gamma($\a_r^0,\b_r^0$), for $r$ and approximating the posterior $p(r|y)$ 
by Gamma($\a_r^q,\b_r^q$), in which the VB parameters $\a_r^q,\b_r^q$ are estimated by the fixed-form VB method of \cite{Salimans:2013}. The fixed-form VB algorithm for updating $\a_r^q,\b_r^q$ is presented in Appendix A.
Empirical Bayes update of $s$ is easier, one can put a Gamma prior on $s$, then the VB optimal posterior of $s$ is also a Gamma. However, we found that, for high-dimensional problems, fixing $s$ to some very small value works better.
We set $s=1e-5$ in our implementation, which implies that we use a very flat prior for the $\l_j$.
We set $S_0=10^4 I$ and $\nu_0=u+1$ in order to have a flat prior on $Q$.

\subsection{Solving \eqref{eq:beta update glmm}}
This section presents a method for solving the optimization problem \eqref{eq:beta update glmm}.
Let 
\beq\label{eq:f(beta)}
f(\b) = [\frac{1}{\phi}]\int \big(1'\zeta(\eta))-y'\eta\big)q(b)db.
\eeq
\eqref{eq:beta update glmm} is equivalent to
\beq\label{eq:beta update glmm1}
\arg\min_\beta\left\{F(\b)=f(\b)+\sum_{j=1}^p[\l_j]|\b_j|\right\}.
\eeq
It's worth noting that the main different between 
\eqref{eq:beta update glmm1} and \eqref{eq:Groll&Schelldorfer}
is that the integral in $f(\b)$ 
can be either computed analytically or approximated easily with an arbitrary accuracy
without relying on the Laplace approximation.
In \eqref{eq:f(beta)} we work with the log-scales of the likelihood,
which is more convenient than with the original scale as in \eqref{eq:Groll&Schelldorfer}.

Recall that $\eta_{ij}=\b_0+x_{ij}'\b_{1:p}+z_{ij}'b_i$ with $b_i\sim \N(\mu_{b_i}^q,\Sig_{b_i}^q)$.
For normal and Poisson regression $\zeta(\eta_{ij})=\eta_{ij}^2$ and $\zeta(\eta_{ij})=e^{\eta_{ij}}$ respectively,
the integral in $f(\b)$ is computed in closed form.
After some algebra, it can be shown that
\beqn
f(\b)=1'\exp\left(X\b+Z\mu_b^q+\frac12\diag(Z\Sigma_b^qZ')\right)-y'(X\b+Z\mu_b^q)
\eeqn
for Poisson regression.
For binomial regression, 
a closed form approximation to $f(\b)$ with an arbitrary accuracy is presented in Appendix B.

That is, the function $f(\b)$ is either computed analytically
or easily approximated with an arbitrary accuracy.
With a little abuse of notation, we still denote the approximation by $f(\b)$ in the latter case.  
An advantage over the method in \cite{Groll:2012} and \cite{Schelldorfer:2013} is that 
our method does not rely on the Laplace approximation for integrating out the random effects.	
The Laplace approximation of the likelihood in GLMMs might be in some cases not very accurate \citep[see, e.g.][]{Joe:2008}.

The optimization problem \eqref{eq:beta update glmm1} 
belongs to a popular class of optimization problems in which the target has the form 
of a sum of a smooth function and a separable convex function.
There are many algorithms available for solving such an optimization problem.
In this paper, we use the coordinate gradient descent method of \cite{Tseng:2009} \cite[see also][]{Schelldorfer:2013} to solve \eqref{eq:beta update glmm1}.

Using the notation in \cite{Schelldorfer:2013}, denote by $\b^{(s)}=(\b_0^{(s)},...,\b_p^{(s)})'$ the value of $\beta$ at the $s$th iteration
and let $\b^{(s,s-1;j)}=(\b_0^{(s)},...,\b_{j-1}^{(s)},\b_{j}^{(s-1)},...,\b_p^{(s-1)})'$.
Let $e_j$ be the $(j+1)$st unit vector and $H^{(s)}_j$ be a positive definite matrix, $j=0,...,p$.
The coordinate gradient descent method is as follows,
whose convergence to a stationary point of $F(\b)$ is proved in \cite{Tseng:2009}.

\begin{enumerate}
\item Initialize $\beta^{(0)}$. Repeat the following for $s=1,2,...$
\item For $j=0,1,...,p$
\begin{itemize}
\item[(i)] Calculate the descent direction
\beq\label{eq: d_j^s}
d_j^{(s)}=\arg\min_d\left\{d\nabla f(\b^{(s,s-1;j)})'e_j+\frac12d^2e_j'H^{(s)}_je_j+[\l_j]|\b_j^{(s-1)}+d|\right\}.
\eeq
\item[(ii)] Choose a step size $\a_j^{(s)}$ and set $\b^{(s,s-1;j+1)}=\b^{(s,s-1;j)}+\a_j^{(s)}d_j^{(s)}e_j$.
\end{itemize}
\end{enumerate}

For matrix $H^{(s)}_j$ we choose $H^{(s)}_j=\nabla^2f(\b^{(s,s-1;j)})$.
It is easy to see that $d_j^{(s)}$ in \eqref{eq: d_j^s} can be solved analytically
\beqn
d_j^{(s)} = 
\begin{cases}
-\frac{\nabla f(\b^{(s,s-1;j)})'e_j}{e_j'H^{(s)}_je_j},&j=0\\
\text{median}\left(\frac{[\l_j]-\nabla f(\b^{(s,s-1;j)})'e_j}{e_j'H^{(s)}_je_j},-\b_j^{(s-1)},\frac{-[\l_j]-\nabla f(\b^{(s,s-1;j)})'e_j}{e_j'H^{(s)}_je_j}\right),&j>0.
\end{cases}
\eeqn
For the step size $\a_j^{(s)}$, \cite{Tseng:2009} suggest the Armijo rule as follows: For some $0<\delta,\varrho<1$ and $0\leq\gamma<1$, choose $\a_j^\text{init}>0$ and let $\a_j^{(s)}$ be the largest element of $\{\a_j^\text{init}\delta^l\}_{l=0,1,...}$ satisfying
\beqn
F(\b^{(s,s-1;j)}+\a_j^{(s)}d_j^{(s)}e_j)\leq F(\b^{(s,s-1;j)})+\a_j^{(s)}\varrho\Delta_j,
\eeqn
where $\Delta_j=d_j^{(s)}\nabla f(\b^{(s,s-1;j)})'e_j+\gamma(d_j^{(s)})^2e_j'H^{(s)}_je_j$ for $j=0$,
and $=d_j^{(s)}\nabla f(\b^{(s,s-1;j)})'e_j+\gamma(d_j^{(s)})^2e_j'H^{(s)}_je_j+[\l_j](|\b_j^{(s-1)}+d_j^{(s)}|-|\b_j^{(s-1)}|)$ for $j>0$.
Following \cite{Schelldorfer:2013}, we choose $\a_j^\text{init}=1,\ \delta=0.5$, $\varrho=0.1$ and $\gamma=0$.

\section{Examples}\label{sec:examples}
\subsection{Simulation study}
We simulate data sets from a mixed effect Poisson regression model
\bean
p(y_{ij}|\b,b_i) &=& \text{Poisson}(\exp(\eta_{ij})),
\eean
and a mixed effect logistic regression model
\bean
p(y_{ij}|\b,b_i) &=& \text{Binomial}\left(\frac{\exp(\eta_{ij})}{1+{\exp(\eta_{ij})}}\right),
\eean
with $\eta_{ij}=\b_0+x_{ij}'\b_{1:p}+z_{ij}'b_i$, $i=1,...,n_i$ and $j=1,...,m$.
Here, $\b_0=3$ and the first four entries of $\b_{1:p}$ are $(-2.5, \ 0,\ 0,\ -2)$ and the rest $p-4$ entries are zeros,
$x_{ij}$ and $z_{ij}$ are independently generated from the uniform distribution on $(0,1)$,
and $b_i\sim \N(0,Q^{-1})$ with $Q=(1/\sigma^2)\mathbb{I}_u$, $n_i$ is set to 5.

We investigate the performance of the proposed VBGLMM approach and compare it to the GLMMLASSO method of \cite{Groll:2012}.
We select the best shrinkage parameter $\l$ in the GLMMLASSO method based on BIC from a range of 100 equally-spaced values between 0 and $\l_{\max}$.
Theoretically, $\l_{\max}$ is the smallest value of $\l$ such that $\b_{1:p}=0$.
Determining $\l_{\max}$ is not straightforward and we set in this simulation example $\l_{\max}=100$ after some experiments.

The performance is measured by the rate of correctly-fitted models (CFR),
mean squared errors in $\beta$ ($\MSE_\beta$), mean squared errors in $\sigma^2$ ($\MSE_{\sigma^2}$),
and CPU time in seconds, over 50 replications.

The simulation results are summarized in Table \ref{tab:poisson} and  Table \ref{tab:binomial} for various scenario with different values of $p$, $m$ and $\sigma^2$.
VBGLMM outperforms GLMMLASSO in all cases. Especially, VBGLMM works very well in terms of identifying correctly the  zero-coefficients.  

\begin{table}
  \begin{center}
    \begin{tabular}{cccccccc}
\hline\hline
$p$	&$m$	&$\sigma^2$	&Method		&CFR(\%)	&$\MSE_\b$	&$\MSE_{\s^2}$	&CPU (seconds)\\
5	&50	&0.5		&glmmlasso	&0		&0.123 		&0.027		&123.7\\ 
	&	&		&vbglmm		&100		&0.091		&0.018		&3.5\\
	&	&1		&glmmlasso	&0		&0.140		&0.031		&278.5\\
	&	&		&vbglmm		&99		&0.101		&0.016		&5.8\\
	&100	&0.5		&glmmlasso	&0		&0.092		&0.024		&377.7\\
	&	&		&vbglmm		&100		&0.079		&0.018		&9.1\\
	&	&1		&glmmlasso	&0 		&0.105		&0.028 		&1491.7\\
	&	&		&vbglmm		&100		&0.092		&0.022 		&32.5 \\
\hline
50	&50	&0.5		&glmmlasso	&0		&1.822		&0.060		&394.7\\
	&	&		&vbglmm		&85		&0.528		&0.035		&17.3\\
	&	&1		&glmmlasso	&0		&1.844		&0.121		&604.6\\
	&	&		&vbglmm		&81		&0.188		&0.051		&19.5\\
	&100	&0.5		&glmmlasso	&0		&0.758		&0.038		&2226.9\\
	&	&		&vbglmm		&89		&0.481		&0.025		&44.3\\
	&	&1		&glmmlasso	&0		&0.738		&0.131		&941.9\\
	&	&		&vbglmm		&82		&0.291		&0.044		&17.4\\
\hline\hline
    \end{tabular}
  \end{center}
  \caption{Simulation: mixed Poisson regression} \label{tab:poisson}
\end{table}

\begin{table}
  \begin{center}
    \begin{tabular}{cccccccc}
\hline\hline
$p$	&$m$	&$\sigma^2$	&Method		&CFR(\%)	&$\MSE_\b$	&$\MSE_{\s^2}$	&CPU (seconds)\\
5	&50	&0.5		&glmmlasso	&0		&1.372		&0.042		&16.3\\ 
	&	&		&vbglmm		&98		&0.580		&0.017		&5.6\\
	&	&1		&glmmlasso	&0		&2.621		&0.469		&20.3\\
	&	&		&vbglmm		&89		&0.675		&0.321		&4.7\\
	&100	&0.5		&glmmlasso	&0		&1.127		&0.055		&63.0\\
	&	&		&vbglmm		&100		&0.541		&0.015		&17.2\\
	&	&1		&glmmlasso	&0		&1.764		&0.521		&102.3\\
	&	&		&vbglmm		&91		&0.656		&0.189		&21.1\\
\hline
50	&50	&0.5		&glmmlasso	&0		&12.408		&0.039		&48.9\\
	&	&		&vbglmm		&72		&1.118		&0.035		&33.2\\
	&	&1		&glmmlasso	&0              &12.300         &0.475          &65.6\\
	&	&		&vbglmm		&72             &1.466          &0.117          &32.1\\
	&100	&0.5		&glmmlasso	&0		&5.306		&0.067		&141.0\\
	&	&		&vbglmm		&74		&0.796		&0.056		&58.8\\
	&	&1		&glmmlasso	&0              &5.281          &0.554          &173.5\\
	&	&		&vbglmm		&80             &1.139          &0.157          &47.1\\
\hline\hline
    \end{tabular}
  \end{center}
  \caption{Simulation: mixed logistic regression} \label{tab:binomial}
\end{table}

\subsection{Skin cancer data}
A clinical trial is conducted to test the effectiveness of beta-carotene in preventing
non-melanoma skin cancer \citep{Greenberg:1989}.
Patients were randomly assigned to a control or treatment group
and biopsied once a year to ascertain the number of new skin cancers since the last examination. 
The response $y_{ij}$ is a count of the number of new skin cancers in year $j$ for the $i$th subject. 
The covariates include \texttt{age, skin} (1 if skin has burns and 0 otherwise), \texttt{gender, exposure} (a count of the number of previous
skin cancers), \texttt{year} of follow-up and \texttt{treatment} (1 if 
the subject is in the treatment group and 0 otherwise). 
There are $m = 1683$ subjects with complete covariate information.

\cite{Donohue:2011} argue that \texttt{treament} is not significant 
and consider 5 different Poisson mixed models
with different inclusion of the rest 5 covariates.
By using an AIC-type model selection criterion,
\cite{Donohue:2011} select a random intercept model with four fixed effect covariates \texttt{age, skin, gender, exposure}
(the fixed effect intercept is always included).

We consider the variable selection problem for this Poisson mixed regression model with a random intercept.
We consider all the 6 potential covariates \texttt{age, skin, gender, exposure, treament} and \texttt{year}.
Our method selects the same model as selected by \cite{Donohue:2011}.
The estimate of vector $\b$ is $(-24.609,\ 0.008,\  0.350,\  1.579,\  0.854,\  0,\ 0)$,
and the estimate of the random effect standard deviation $\s$ is $102.7$.

\subsection{Six city data}
The six cities dataset in \cite{Fitzmaurice:1993} consists of binary responses $y_{ij}$ which indicate the wheezing status (1 if wheezing, 0
if not wheezing) of the $i$th child at time-point $j$, $i = 1, . . . , 537$ and $j = 1,...,4$.
The covariates are \texttt{Age} (the age of the child at time-point $j$, centered at 9 years) and \texttt{Smoke} (the maternal smoking status 0 or 1).
We consider the following logistic mixed regression model with two random effects
\begin{eqnarray*}
p(y_{ij}|\beta,b_i)&=&\text{Binomial}(1,p_{ij}),\\
\text{logit}(p_{ij})&=&\beta_0+\beta_1\text{Age}_{ij}+\beta_2\text{Smoke}_{ij}+b_{i1}+b_{i2}\text{Age}_{ij}.
\end{eqnarray*}
The VBGLMM estimate of $\b$ is $(-6.98,\ 0,\ 0)$,
i.e. \texttt{Age} and \texttt{Smoke} are not selected.
The estimate of the covariance matrix of the random effects $b_i$ is 
\beqn
\wh{\text{Cov}}(b_i)=\begin{pmatrix}
34.863 & -1.103\\
-1.103 & 0.434
\end{pmatrix}.
\eeqn

\section{Conclusions and Discussions}\label{sec:conclusion}
We have described in this article a VB algorithm for simultaneous variable selection 
and parameter estimation in GLMMs.
The proposed algorithm is based on the VB method for estimating a posterior mode
in conjunction with the Bayesian adaptive Lasso.
The posterior mode VB method described in this article 
can be applied to variable selection in other frameworks such as covariance selection.
The proposed VBGLMM method can also be extended 
to (i) grouped variable selection in GLMMs by using the group lasso penalty \citep{Yuan:2006}
(ii) ordered variable selection in GLMMs by the composite absolute penalty \citep{Zhao:2009}.
This research is currently in progress.

\subsection*{Appendix A: Fixed-form VB algorithm for approximating $p(r|y)$}
This section presents the fixed-form VB approach of \cite{Salimans:2013}
for approximating $p(r|y)$. Their fixed-form VB algorithm requires an unbiased estimate of a covariance matrix of the form $\cov(T(X),V(X))$
with $T(\cdot)$ and $V(\cdot)$ vector functions of a random variable $X$ with probability density function $f(x)$.
Let $X_1$ and $X_2$ be two independent draws from $f$.
It is easy to see that
\beqn
\wh\cov=\frac12 (T(X_1)-T(X_2))(V(X_1)-V(X_2))'
\eeqn
is an unbiased estimate of $\cov(T(X),V(X))$.

We use a Gamma prior Gamma($\a_r^0,\b_r^0$) for $r$ and approximate the posterior $p(r|y)$ 
by $q(r)=\text{Gamma}(\a_r^q,\b_r^q)$.
The sufficient statistic for the natural parameter $\eta=(\a_r^q,\b_r^q)'$ is $T(r)=(\log r,-r)'$ and
\beqn
\log p(r,y)=\left(p\log s-\beta_r^0+\sum_{j=1}^p[\log\l_j]\right)r+(\alpha_r^0-1)\log r-p\log\Gamma(r)
\eeqn
after ignoring the terms independent of $r$.
Let 
\beqn
C=C(\a_r^q,\b_r^q)=\begin{pmatrix}
\dot\psi(\a_r^q) &-\frac{1}{\b_r^q}\\
-\frac{1}{\b_r^q}&\frac{\a_r^q}{{\b_r^q}^2}
\end{pmatrix}.
\eeqn
We have the following algorithm for estimating $\a_r^q$ and $\b_r^q$.
\begin{enumerate}
\item Initialize $\eta=(\a_r^q,\b_r^q)'$. Compute $C=C(\a_r^q,\b_r^q)$ and $g=C\eta$.
\item Initialize $\bar C=0$, $\bar g=0$.
\item For $i=1,2,...,N$
\begin{itemize}
\item Set $\eta=C^{-1}g$
\item Generate $r_1,r_2$ from $q(r)$ and compute
\beqn
\wh g_i=\frac12 (\log p(r_1,y)-\log p(r_2,y))(T(r_1)-T(r_2))
\eeqn
and $\wh C_i=C(\a_r^q,\b_r^q)$.
\item Set $g=(1-c)g+c\wh g_i$, $C=(1-c)C+c\wh C_i$.
\item If $i>N/2$ set $\bar g=\bar g+\wh g_i$, $\bar C=\bar C+\wh C_i$.
\end{itemize}
\item Set $\eta={\bar C}^{-1}\bar g$.
\end{enumerate}

\subsection*{Appendix B}
For binomial mixed regression, $\zeta(\eta_{ij})=\log(1+e^{\eta_{ij}})$,
where $\eta_{ij}$ is normally distributed with mean $\b_0+x_{ij}'\b_{1:p}+z_{ij}'\mu_{b_i}^q$  and variance $z_{ij}'\Sig_{b_i}^qz_{ij}$.
The function $f(\b)$ in \eqref{eq:f(beta)} becomes
\beqn
f(\b)=\sum_{i,j}\E_{\eta_{ij}}(\log(1+e^{\eta_{ij}}))-y'(\b_0+X\b_{1:p}+Z\mu_b^q).
\eeqn
Computing $f(\b)$ reduces to computing the integrals of the from $\E_\xi(\log(1+e^\xi))$ with $\xi\sim \N(\mu,\s^2)$.
We write $\E_\xi(\log(1+e^\xi))=\E_\zeta h(\zeta)$ with $h(\zeta)=\log(1+e^{\mu+\s \zeta})$ and $\zeta\sim \N(0,1)$.
Using the Taylor expansion of $h(\zeta)$ at zero, $h(\zeta)$ can be approximated by
\beqn
h(\zeta) \approx h(0)+\sum_{k=1}^K\frac{h^{(k)}(0)}{k!}\zeta^k
\eeqn
for some $K\geq1$.
Hence, 
\beqn
\E_\xi(\log(1+e^\xi))\approx h(0)+\sum_{k=1}^K\frac{h^{(k)}(0)}{k!}\E_\zeta(\zeta^k).
\eeqn
Note that $\E_\zeta(\zeta^{k})=0$ if $k$ is odd and $\E_\zeta(\zeta^{k})=(k-1)!!$ 
if $k$ is even, where $(k-1)!!=1.3...(k-1)$, i.e. the product of every odd number from 1 to $k-1$.
We set $K=2$ in the examples reported in this article.
The user can set a bigger $K$ in the \texttt{R} package \texttt{vbglmm}.

\subsection*{Appendix C: Gaussian approximation}
Suppose that $q(x)=e^{f(x)}$ and we wish to approximate $q(x)$ by a Gaussian density.
Let $x^*$ be the maximizer of $f(x)$. By Taylor's expansion
\[f(x)\approx f(x^*)+\frac12(x-x^*)'\frac{\partial^2 f(x^*)}{\partial x\partial x'}(x-x^*).\] 
Then
\bean
q(x)=e^{f(x)}&\approx&\exp\big(f(x^*)+\frac12(x-x^*)'\frac{\partial^2 f(x^*)}{\partial x\partial x'}(x-x^*)\big)\\
&\propto&\exp\big(\frac12(x-x^*)'\frac{\partial^2 f(x^*)}{\partial x\partial x'}(x-x^*)\big).
\eean
So the best Gaussian approximation to $q(x)$ has mean $x^*$ and covariance matrix $-(\frac{\partial^2 f(x^*)}{\partial x\partial x'})^{-1}$.

Recall that we wish to maximize 
\[h(b)=-\frac12b'[Q_b]b+[\frac{1}{\phi}](y'\eta-1'\zeta(\eta)),\]
with $\eta=X\beta^q+Zb$. 
The first and second derivatives are
\bean
u(b)=\frac{\partial h(b)}{\partial b}&=& [\frac{1}{\phi}]Z'(y-\dot{\zeta}(\eta))-[Q_b]b\\
H(b)=\frac{\partial^2 h(b)}{\partial b\partial b'} &=& -[\frac{1}{\phi}]Z' \mbox{diag}\left(\ddot{\zeta}(\eta)\right)Z - [Q_b].
\eean
The Newton-Raphson method for maximizing $h(b)$: 
\begin{enumerate}
\item Initialize $b^\text{old}$.
\item Update until some stopping rule is satisfied
\[b^\text{new}=b^\text{old}-H(b^\text{old})^{-1}u(b^\text{old}).\]
\end{enumerate}

\bibliographystyle{apalike}
\bibliography{references_v1}

\begin{thebibliography}{}

\bibitem[Dempster et~al., 1977]{Dempster:1977}
Dempster, A., Laird, N., and Rubin, D. (1977).
\newblock Maximum likelihood from incomplete data via the {EM} algorithm.
\newblock {\em Journal of the Royal Statistical Society, Series B},
  39(1):1--38.

\bibitem[Donohue et~al., 2011]{Donohue:2011}
Donohue, M.~C., Overholser, R., Xu, R., and Vaida, F. (2011).
\newblock Conditional {A}kaike information under generalized linear and
  proportional hazards mixed models.
\newblock {\em Biometrika}, 98:685--700.

\bibitem[Fitzmaurice and Laird, 1993]{Fitzmaurice:1993}
Fitzmaurice, G. and Laird, N. (1993).
\newblock A likelihood-based method for analysing longitudinal binary
  responses.
\newblock {\em Biometrika}, 80:141--151.

\bibitem[Greenberg et~al., 1989]{Greenberg:1989}
Greenberg, E.~R., Baron, J.~A., Stevens, M.~M., Stukel, T.~A., Mandel, J.~S.,
  Spencer, S.~K., Elias, P.~M., Lowe, N., Nierenberg, D.~N., G., B., and Vance,
  J.~C. (1989).
\newblock The skin cancer prevention study: design of a clinical trial of
  beta-carotene among persons at high risk for nonmelanoma skin cancer.
\newblock {\em Controlled Clinical Trials}, 10:153--166.

\bibitem[Griffin and Brown, 2011]{Griffin:2011}
Griffin, J.~E. and Brown, P.~J. (2011).
\newblock Bayesian adaptive {Lassos} with non-convex penalization.
\newblock {\em Australian and New Zealand Journal of Statistics}, 53:423--442.

\bibitem[Groll and Tutz, 2012]{Groll:2012}
Groll, A. and Tutz, G. (2012).
\newblock Variable selection for generalized linear mixed models by
  l1-penalized estimation.
\newblock {\em Statistics and Computing}, pages 1--18.

\bibitem[Joe, 2008]{Joe:2008}
Joe, H. (2008).
\newblock Accuracy of {Laplace} approximation for discrete response mixed
  models.
\newblock {\em Computational Statistics \& Data Analysis}, 52(12):5066 -- 5074.

\bibitem[Leng et~al., 2013]{Leng:2013}
Leng, C., Tran, M.-N., and Nott, D.~J. (2013).
\newblock Bayesian adaptive lasso.
\newblock {\em The Annals of the Institute of Statistical Mathematics}.
\newblock To appear.

\bibitem[Park and Casella, 2008]{Park:2008}
Park, T. and Casella, G. (2008).
\newblock The bayesian lasso.
\newblock {\em Journal of the American Statistical Association}, 103:681--686.

\bibitem[Salimans and Knowles, 2013]{Salimans:2013}
Salimans, T. and Knowles, D.~A. (2013).
\newblock Fixed-form variational posterior approximation through stochastic
  linear regression.
\newblock Technical report, Erasmus University Rotterdam.
\newblock Available at http://arxiv.org/abs/1206.6679.

\bibitem[Schelldorfer et~al., 2013]{Schelldorfer:2013}
Schelldorfer, J., Meier, L., and Bühlmann, P. (2013).
\newblock {GLMMLasso}: An algorithm for high-dimensional generalized linear
  mixed models using l1-penalization.
\newblock {\em Journal of Computational and Graphical Statistics}, 0(ja):null.

\bibitem[Tibshirani, 1996]{Tibshirani:96}
Tibshirani, R. (1996).
\newblock Regression shrinkage and selection via the lasso.
\newblock {\em Journal of the Royal Statistical Society B}, 58(1):267--288.

\bibitem[Tseng and Yun, 2009]{Tseng:2009}
Tseng, P. and Yun, S. (2009).
\newblock A coordinate gradient descent method for nonsmooth separable
  minimization.
\newblock {\em Mathematical Programming}, 117(1-2):387--423.

\bibitem[Yuan and Lin, 2006]{Yuan:2006}
Yuan, M. and Lin, Y. (2006).
\newblock Model selection and estimation in regression with grouped variables.
\newblock {\em Journal of the Royal Statistical Society, Series B}, 68:49--67.

\bibitem[Zhao et~al., 2009]{Zhao:2009}
Zhao, P., Rocha, G., and Yu, B. (2009).
\newblock The composite absolute penalties family for grouped and hierarchical
  variable selection.
\newblock {\em The Annals of Statistics}, 37:3468–3497.

\bibitem[Zou, 2006]{Zou:2006}
Zou, H. (2006).
\newblock The adaptive {Lasso} and its oracle properties.
\newblock {\em Journal of the American Statistical Association},
  101(476):1418--1429.

\end{thebibliography}

\end{document}